\documentclass[pdflatex,sn-mathphys,Numbered,referee]{sn-jnl}

\usepackage{graphicx}%
\usepackage{multirow}%
\usepackage{amsmath,amssymb,amsfonts}%
\usepackage{amsthm}%
\usepackage{mathrsfs}%
\usepackage[title]{appendix}%
\usepackage{xcolor}%
\usepackage{textcomp}%
\usepackage{manyfoot}%
\usepackage{booktabs}%
\usepackage{algorithm}%
\usepackage{algorithmicx}%
\usepackage{algpseudocode}%
\usepackage{listings}%
\usepackage{bm}
\usepackage{comment}

\theoremstyle{thmstyleone}%
\theoremstyle{thmstyletwo}%

\theoremstyle{thmstylethree}%

\def\omit#1{}

\def\V#1{\bm{#1}}

\begin{document}

\title[Andreev reflection contribution to surface-electron mobility]
{Contribution of Andreev reflection to the mobility of
surface state electrons on superfluid $^3$He-B}


\author*[1]{\fnm{Yasumasa} \sur{Tsutsumi}}\email{y.tsutsumi@kwansei.ac.jp}

\author[2]{\fnm{Hiroki} \sur{Ikegami}}\email{hikegami@iphy.ac.cn}

\author[3]{\fnm{Kimitoshi} \sur{Kono}}\email{kkono@nycu.edu.tw}

\affil*[1]{\orgdiv{Department of Physics}, \orgname{Kwansei Gakuin University}, \orgaddress{\city{Sanda}, \postcode{669-1330}, \state{Hyogo}, \country{Japan}}}

\affil[2]{\orgdiv{Beijing National Laboratory for Condensed Matter Physics},
  \orgname{Institute of Physics, Chinese Academy of Sciences},
  \orgaddress{\city{Beijing}, \postcode{100190},
    \country{China}}}

\affil[3]{\orgdiv{Department of Electrophysics},
  \orgname{National Yang Ming Chiao Tung University (NYCU)},
  \orgaddress{\city{Hsinchu}, \postcode{300093},
    \country{Taiwan}}}


\abstract{
The mobility of the Wigner solid on the superfluid $^3$He is determined by the momentum transfer from the scattered $^3$He quasiparticles at the free surface.
The scattering process of the quasiparticles is classified into the normal reflection and the Andreev retroreflection.
Since the quasiparticles nearly conserve the momentum in the process of the Andreev retroreflection at the free surface, the Andreev reflected quasiparticles do not produce a resistive force to the Wigner solid.
In this report, we have analytically calculated the contribution of the Andreev retroreflection to the mobility of the Wigner solid on superfluid $^3$He-B by employing a realistic model order parameter with the free surface.
The Andreev retroreflection is lacked for quasiparticles with energy above the bulk energy gap under the model order parameter.
Then, the Andreev retroreflection does not contribute to a rise in the mobility of the Wigner solid on the superfluid $^3$He-B.
The present model calculation is in good agreement with the previous experimental observation.
We have also discussed the Andreev retroreflection under a self-consistently calculated order parameter.
}


\maketitle

\section{Purpose}\label{sec1}

Andreev suggested an unusual scattering process of quasiparticle from
the region where the order parameter varies in
space~\cite{Andreev1964}.  It successfully interpreted the thermal
conductivity of superconductors.  Andreev's unconventional scattering
interprets phenomena not only in superconductors but also in
superfluid
$^3$He~\cite{Leggett1975,Wheatley1975,Halperin1990,Kurkijarvi1990}.
Thermal boundary resistance of superfluid $^3$He-B was studied, where
the exponential temperature dependence was discussed along with the
Andreev scattering~\cite{Parpia1985}.  The slip length of shear
viscosity is explained by the Andreev
scattering~\cite{Einzel1983,Einzel1984,Einzel1987,Matsubara1999}.  The
motion of an object in superfluid $^3$He is subject to the Andreev
scattering~\cite{Guenault1986,Fisher1989,Bradley2008,Defoort2016}.
The Andreev scattering is also studied in the context of quantum
turbulence in superfluid
$^3$He~\cite{Fisher2014,Bradley2017,Tsepelin2017}.  The direct
observation of Andreev scattering from a moving paddle or from the
free surface of superfluid $^3$He is
reported~\cite{Enrico1993,Okuda1998}, in which quasiparticle beam
emitted from the so-called black-body radiator is employed.  It is
pointed out that quasiparticle scattering by ions can also be viewed
as a special form of Andreev scattering, because it is accompanied by
branch conversion~\cite{Kurkijarvi1990}.  This is beautifully
demonstrated in the recent
studies~\cite{Ikegami2013,Ikegami2013b,Ikegami2015,%
  Shevtsov2016,Tsutsumi2017,Ikegami2019}.

In the above-mentioned examples of Andreev scattering in superfluid
$^3$He, except for the ion case, the agreement between experiment and
theory remains qualitative.  This is either due to a difficulty in
specifying the properties of interface between superfluid $^3$He and
the object or due to the ambiguity in the experimental
geometry~\cite{Okuda1998}.  To eliminate these ambiguities, the
mobility measurement of Wigner solid on the surface of superfluid
$^3$He-B should be one of the most promising experimental methods to
quantitatively verify the Andreev
scattering~\cite{Grimes1979,Fisher1979,Shirahama1997,Shirahama1998,%
  Ikegami2006, Kono2008}.  Looking back at our previous
data~\cite{Shirahama1997}, the resistivity (normalized at superfluid
transition temperature) of the Wigner solid on superfluid $^3$He-B is
satisfactorily fit by the following formula:
\begin{equation}
  \label{eq:fermi-function}
  \frac{R(T)}{R(T_c)}=\frac{2}{e^{\Delta(T)/k_BT}+1},
\end{equation}
where $T_c$ is the superfluid transition temperature at the saturated
vapor pressure (930~$\mu$K), $\Delta(T)$ the superfluid energy gap of
B phase, $k_B$ the Boltzmann constant.  This implies that the normal
quasiparticle scattering off from a specular free surface is
predominant.  But, if we look at the data (Fig. 4 of
Ref.~\cite{Shirahama1997}), downward deviation of the resistance from
Eq.~(\ref{eq:fermi-function}) may be seen, which may be an indication
of the Andreev scattering contribution.  Incidentally, in that work,
an unresolved pressing electric field dependence exists, which is not
reproduced in a more recent work~\cite{Ikegami2006}.  The temperature
may not yet be sufficiently low to observe a significant contribution
from the Andreev scattering.  
In this report, we analytically calculate the contribution of the Andreev retroreflection to the mobility of the Wigner solid on superfluid $^3$He-B by employing a realistic model order parameter with the free surface.  The model calculation is in good agreement with the previous experimental observation.

\section{Methods}\label{sec2}

The theoretical analysis of the mobility of the Wigner solid on
superfluid $^3$He is developed by Monarkha and one of the present
authors~\cite{Monarkha1997,Monarkha2004,Kono2008}.  In that theory,
only the normal scattering of quasiparticle from the free surface is
taken into account.  As shown below, the implementation of Andreev
scattering should be straightforward, at least formally.

\begin{figure}
  \begin{center} \resizebox{0.7\textwidth}{!}{%
\includegraphics{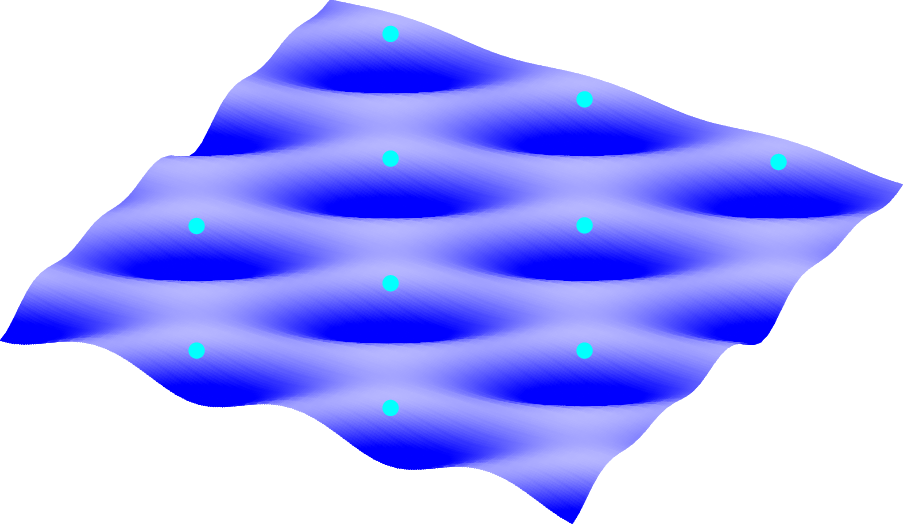}}
\caption{\label{fig:dimple} (Color online) Schematic illustration of a
  dimple lattice formed under the Wigner solid.  When 2D electrons
  solidify to form a Wigner solid, electron starts to localize at each
  lattice site, so that the diffusion speed becomes sufficiently slow
  for the surface to respond to the force exerted by electrons, which
  forms a triangular lattice.}
  \end{center}
\end{figure}

The Wigner solid on the surface of liquid helium forms a dimple
lattice on the surface~\cite{Fisher1979,Monarkha1997,Monarkha1998},
which is schematically illustrated in Fig.~\ref{fig:dimple}.
Solidification results in a significant slowing down of electron
diffusion velocity, so that helium surface could respond to nonuniform
electron pressure localized at each lattice site.  When the Wigner
solid is driven parallel to the average plane of the surface, the
Wigner solid moves accompanied by the dimple lattice.  A fluid
velocity gradient induced by the motion of dimple lattice results in a
resistive force to the Wigner solid due to the viscosity at high
temperatures.  This is observed in the normal phase of liquid
$^3$He~\cite{Suto1996,Monarkha1998,Kono2002,Kono2010}.  The fluid
velocity gradient exists below the surface only within a depth of
lattice constant of the Wigner solid.  Below 10~mK a quasiparticle
mean-free-path length exceeds this length scale, and the system enters
a ballistic regime.  In this regime, the temperature dependence
disappears.  When the qasiparticle is reflected by the free surface, a
momentum change (impulse) of quasiparticle before and after the
scattering is transferred to the surface and resulting force is
balanced with the external pressure.  Here, only the electron pressure
associated with a lateral motion is of interest to account for the
electric resistance.  A stationary Wigner solid with respect to the
sample cell results in no lateral force because quasiparticles
equilibrate with the wall of sample cell and impulse due to
quasiparticle reflection should cancel with each other.  The drag
force to resist the lateral motion of the Wigner solid can be expanded
in terms of the drift velocity $\V{v}_d$ of Wigner
solid~\cite{Monarkha1997,Kono2008}.  The drag force first order in
$\V{v}_d$ (Eq.(6) of Ref.~\cite{Kono2008}) is given by
\begin{equation}
  \label{eq:drag-force}
  \V{F}_t=-2\V{v}_d\int_C{}dS\,(\V{n}\cdot\hat{\V{v}}_d)^2%
  \sum_{\sigma}\int{}d\Omega'\int_0^{\infty}{}\frac{p^2}{(2\pi\hbar)^3}dp\ %
  (\V{n}\cdot\V{p})^3\frac{1}{p}%
  \frac{\partial\epsilon_{p\sigma}}{\partial{}p}%
  \left(-\frac{\partial{}f}{\partial\epsilon_{p\sigma}}\right),
\end{equation}
where $\hat{\V{v}}_d$ is a unit vector parallel to $\V{v}_d$; $dS$ and
$\V{n}$ denote a surface element and surface normal of liquid He
surface, respectively, and $C$ is the surface region of interest;
$\V{p}$ and $\sigma$ denote a momentum and spin of quasiparticle,
respectively, and $\epsilon_{p\sigma}$ is its energy; $\int{}d\Omega'$
indicates integration over the solid angle which satisfies the
condition $(\V{n}\cdot\V{p})\ge0$, implying that quasiparticles come
only from the liquid side; $f(x)$ is the Fermi distribution function,
\[
  f(x)=\frac{1}{e^{(x-\mu)/k_BT}+1},
\]
where $\mu$ is the chemical potential and $k_B$ is the Boltzmann
constant.  Inside the solid angle and
momentum integral, $(\V{n}\cdot\V{p})$ may be replaced by
$(\overline{\V{n}}\cdot\V{p})$ because the dimple is actually very
shallow, where $\overline{\V{n}}$ is an average value of $\V{n}$:
$(\overline{\V{n}}\cdot\V{v}_d)=0$, corresponding to a flat plane of
liquid surface without dimples.  Because
$\partial{}f/\partial\epsilon$ is sharply peaked at around the chemical
potential value or the Fermi energy, we may set $p$ to the Fermi
momentum $p_F$ and take it outside integral, thus,
Eq.~(\ref{eq:drag-force}) is simplified as
\begin{equation}
  \label{eq:drag-force-2}
  \V{F}_t=-2\V{v}_d\left(\int_C{}dS(\V{n}\cdot\hat{\V{v}}_d)^2\right)%
  \frac{p_F^4}{4\pi^2\hbar^3}\sum_{\sigma}\int_0^{\pi/2}d\theta%
  \int_0^{\infty}{}dp\ %
  \cos^3\theta\sin\theta\,\,%
  \frac{\partial\epsilon_{p\sigma}}{\partial{}p}%
  \left(-\frac{\partial{}f}{\partial\epsilon_{p\sigma}}\right).
\end{equation}
Here $\theta$ is a polar angle from $\overline{\V{n}}$ and axial
symmetry around $\overline{\V{n}}$ is used.  The drift velocity is
expressed by using the mobility of Wigner solid $\mu_w$ as
$\V{v}_d=-\mu_w\V{E}_{\parallel}$, and the resistive drag force should
be balanced with the electric force on the Wigner solid $-N_se\V{E}$,
so that we obtain the expression for $\mu_w$
\[
  \frac{e}{\mu_w}=\frac{2}{N_s}%
  \left(\int_C{}dS(\V{n}\cdot\hat{\V{v}}_d)^2\right)%
  \frac{p_F^4}{4\pi^2\hbar^3}\sum_{\sigma}\int_0^{\pi/2}d\theta
  \int_0^{\infty}dp\ %
  \cos^3\theta\sin\theta\,\,%
  \frac{\partial\epsilon_{p\sigma}}{\partial{}p}%
  \left(-\frac{\partial{}f}{\partial\epsilon_{p\sigma}}\right).
\]
Here, $\V{E}_{\parallel}$ is a driving electric field applied parallel
to the liquid He surface, $N_s$ a number of electrons in the region $C$,
and $e$ is the elementary charge.  In the normal liquid $^3$He, $f(x)$
abruptly varies from 1 to 0 around $\epsilon_{p\sigma}=\mu$, and
hence, the Wigner solid mobility on normal liquid $^3$He $\mu_w^n$ is
given by
\begin{equation}
  \frac{e}{\mu_w^n}=\frac{p_F^4}{4\pi^2\hbar^3N_s}%
  \left(\int_C{}dS(\V{n}\cdot\hat{\V{v}}_d)^2\right).
\end{equation}
The integral in parentheses is determined only by the profile of surface dimple
lattice~\cite{Monarkha1997,Monarkha1998}.
In the following, we deal with the normalized mobility
\begin{equation}
  \label{eq:normalized-mobility}
  \frac{\mu_w^n}{\mu_w}=2\sum_{\sigma}\int_0^{\pi/2}d\theta\int_0^{\infty}dp\ %
  \cos^3\theta\sin\theta\,\,%
  \frac{\partial\epsilon_{p\sigma}}{\partial{}p}%
  \left(-\frac{\partial{}f}{\partial\epsilon_{p\sigma}}\right).
\end{equation}
\begin{figure}
  \begin{tabular}{cc}
  \begin{minipage}[c]{0.5\textwidth}
    \centering
    \resizebox{0.8\textwidth}{!}{\includegraphics{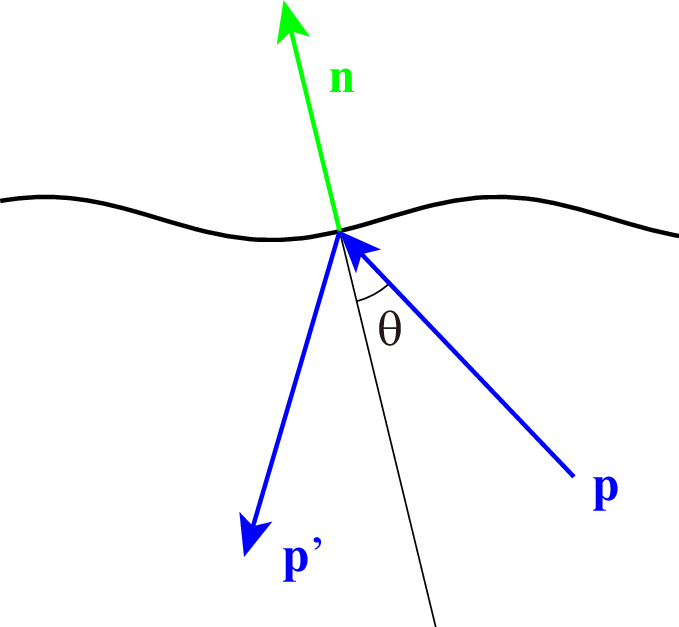}}
  \end{minipage}
  &
  \begin{minipage}[c]{0.5\textwidth}
    \centering
      \resizebox{0.8\textwidth}{!}{\includegraphics{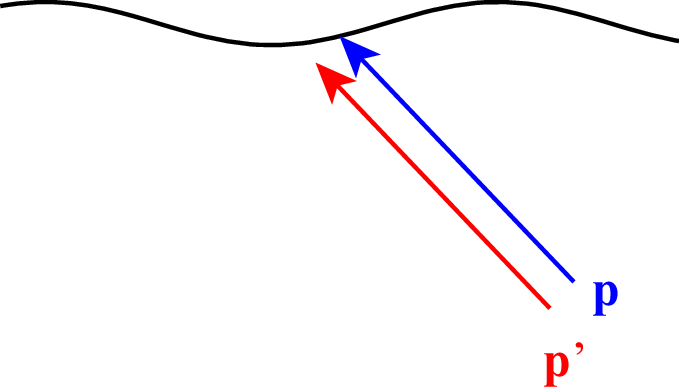}}
  \end{minipage}\\
    (a) & (b)
  \end{tabular}\\[6pt]
  \caption{\label{fig:scattering} (Color online).  Schematic diagrams of the
    momentum of scattered quasiparticle (hole for the Andreev
    scattering case).  A ballistically impinging quasiparticle of
    momentum $\V{p}$ is scattered either into a quasiparticle of
    momentum $\V{p}'$ (a) or into a quasihole of momentum $\V{p}'=\V{p}$ (b)
    at the free surface of superfluid $^3$He.  (a) A impinging
    quasiparticle is specularly scattered by a local tangential plane
    of the surface.  The momentum vector of the scattered particle is
    parallel to the trajectory of the quasiparticle.  The momentum
    projection to the surface normal is inverted, while the momentum
    in the tangential place preserves in the normal scattering
    case. (b) As for the Andreev scattering case, branch conversion
    takes place and the momentum of scattered quasihole is nearly
    identical to that of impinging quasiparticle.  The velocity of
    quasihole is opposite to the momentum $\V{p}'$, and the hole
    travels back in the same trajectory of impinging quasiparticle.
    No momentum change results and there is no momentum transfer to
    the surface is anticipated.}
\end{figure}%

The argument so far only takes into account normal reflection of
quasiparticle from the surface, which is illustrated in
Fig.~\ref{fig:scattering}a.  In the superfluid phase, however, in
addition to the normal reflection, so-called Andreev retroreflection
exists.  In the Andreev retroreflection, an incident quasiparticle
(quasihole) is retroreflected as a quasihole (quasiparticle) with
opposite velocity and nearly equal momentum to the incident
quasiparticle (quasihole), as demonstrated in
Fig.~\ref{fig:scattering}b.  In this case, because no momentum change
exists, the scattering event does not produce an impulse to the
surface. 

Therefore, this part should be subtracted from the above
expression Eq.~(\ref{eq:normalized-mobility}).  Andreev reflection
coefficients at the free surface of superfluid $^3$He, which is
specular, are calculated for a fixed order
parameter~\cite{Kurkijarvi1990,Kieselmann1983,Buchholtz1981}, or for a
self-consistently determined order parameter~\cite{Nagato1998}.
Following the notation of Ref.~\cite{Kurkijarvi1990,Kieselmann1983},
the normal reflection and Andreev retroreflection coefficients are
denoted by $T_{00}$ and $R_{00}$, respectively.  These coefficients satisfy the relation $T_{00}+R_{00}=1$.  The integrand in
Eq.~(\ref{eq:normalized-mobility}) must be multiplied by $T_{00}$.  By
introducing a new variable $\eta_p=v_F(p-p_F)$ substituting $p$, and
measure quasiparticle energy from the chemical potential or Fermi
energy, an appropriate expression for the normalized mobility should
be
\begin{equation}
  \label{eq:normalized-mobility-2}
  \frac{\mu^n}{\mu}=8\int_0^{\pi/2}d\theta\int_0^{\infty}d\eta_p\ %
  \cos^3\theta\sin\theta\,\,%
  T_{00}(\epsilon(\eta_p),\theta)
  \left|\frac{\partial\epsilon(\eta_p)}{\partial{}\eta_p}\right|%
  \left(-\frac{\partial{}f}{\partial\epsilon(\eta_p)}\right),
\end{equation}
where suffix $w$ is omitted for simplicity,
quasiparticle energy $\epsilon(\eta_p)=\sqrt{\eta_p^2+\Delta^2}$ is independent of the spin state
with a superfluid energy gap $\Delta$, and
$f(\epsilon)=[\exp(\epsilon/k_BT)+1]^{-1}$. 


The normal reflection and Andreev retororeflection coefficients are obtained by the boundary condition at the free surface, in which the superposition wave function of the injected quasiparticle and the reflected quasiparticle vanishes.
We can find quasiparticle wave functions analytically from the Andreev equation under a realistic model order parameter with the free surface.
The following section shows the analytical solutions of the Andreev equation and the comparison of the mobility of the Wigner solid between the calculated mobility from Eq.~\eqref{eq:normalized-mobility-2} and the observed mobility in the previous experiment~\cite{Ikegami2006}.

\section{Results}\label{sec3}

\subsection{Analytical solutions of Andreev equation}

First, we find analytical solutions of the Andreev equation giving the quasiparticle wave function:
\begin{equation}
\begin{pmatrix}
-i\alpha\hbar v_{\perp}\partial_z\hat{\sigma}_0 & \hat{\Delta}_{\alpha}(z,{\bm p}_{\parallel}) \\
\hat{\Delta}_{\alpha}^{\dagger}(z,{\bm p}_{\parallel}) & i\alpha\hbar v_{\perp}\partial_z\hat{\sigma}_0
\end{pmatrix}
\phi^{\alpha}(z,{\bm p}_{\parallel})=\epsilon\phi^{\alpha}(z,{\bm p}_{\parallel}),
\end{equation}
with the order parameter
\begin{equation}
\hat{\Delta}_{\alpha}(z,{\bm p}_{\parallel})=
\begin{pmatrix}
-\Delta_{\parallel}(z)\sin\theta e^{-i\phi} & \alpha\Delta_{\perp}(z)\cos\theta \\
\alpha\Delta_{\perp}(z)\cos\theta & \Delta_{\parallel}(z)\sin\theta e^{i\phi}
\end{pmatrix}.
\end{equation}
We take the $z$-axis normal to the free surface and ${\bm p}_{\parallel}=p_{\rm F}\sin\theta(\cos\phi,\sin\phi)$ indicates components of the Fermi momentum parallel to the free surface.
The perpendicular component of the Fermi momentum is described by $p_{\perp}=\alpha\sqrt{p_{\rm F}^2-{\bm p}_{\parallel}^2}=\alpha p_{\rm F}\cos\theta$ with the sign $\alpha$ and $0\le\theta\le\pi/2$.
Here, $\hat{\sigma}_0$ is the unit matrix in the spin space and $v_{\perp}=v_{\rm F}\cos\theta$ indicates perpendicular component of the Fermi velocity.

We assume that the free surface is situated at $z=0$ and the $^3$He-B fills the region $z>0$.
In this report, we use a model potential $\Delta_{\parallel}(z)=\Delta$ and $\Delta_{\perp}(z)=\Delta\tanh\tilde{z}$, where we define a dimensionless length $\tilde{z}\equiv z/\xi$ with the coherence length $\xi=\hbar v_{\rm F}/\Delta$.
Note that the model potential is the self-consistent pair potential in the weak coupling limit~\cite{Tsutsumi2012}.
We can solve the Andreev equation analytically under this model potential.
The four solutions of the wave function for the quasiparticle energy $\epsilon$ are obtained as
\begin{equation}
\begin{split}
&\phi_{\rm p\uparrow}^{\alpha}(z,{\bm p}_{\parallel})=\exp\left[i\alpha\frac{\Omega}{\hbar v_{\rm \perp}}z\right]
\begin{pmatrix}
u^{\alpha}(z,\theta) \\ -e^{i\phi}w^{\alpha}(z,\theta) \\ -e^{i\phi}v_{\parallel}^{\alpha}(z,\theta) \\ v_{\perp}^{\alpha}(z,\theta) 
\end{pmatrix},\
\phi_{\rm p\downarrow}^{\alpha}(z,{\bm p}_{\parallel})=\exp\left[i\alpha\frac{\Omega}{\hbar v_{\rm \perp}}z\right]
\begin{pmatrix}
e^{-i\phi}w^{\alpha}(z,\theta) \\ u^{\alpha}(z,\theta)  \\ v_{\perp}^{\alpha}(z,\theta) \\ e^{-i\phi}v_{\parallel}^{\alpha}(z,\theta)
\end{pmatrix},\\
&\phi_{\rm h\uparrow}^{\alpha}(z,{\bm p}_{\parallel})=\exp\left[-i\alpha\frac{\Omega}{\hbar v_{\rm \perp}}z\right]
\begin{pmatrix}
-e^{-i\phi}v_{\parallel}^{-\alpha}(z,\theta) \\ -v_{\perp}^{-\alpha}(z,\theta) \\ u^{-\alpha}(z,\theta) \\ e^{-i\phi}w^{-\alpha}(z,\theta)
\end{pmatrix},\
\phi_{\rm h\downarrow}^{\alpha}(z,{\bm p}_{\parallel})=\exp\left[-i\alpha\frac{\Omega}{\hbar v_{\rm \perp}}z\right]
\begin{pmatrix}
-v_{\perp}^{-\alpha}(z,\theta) \\ e^{i\phi}v_{\parallel}^{-\alpha}(z,\theta) \\ -e^{i\phi}w^{-\alpha}(z,\theta) \\  u^{-\alpha}(z,\theta)
\end{pmatrix},
\end{split}
\end{equation}
where $\phi_{\rm p\sigma}^{\alpha}$ and $\phi_{\rm h\sigma}^{\alpha}$ indicate particle-like and hole-like quasiparticles, respectively, with the spin $\sigma$.
The spatial variation of each component is given by
\begin{equation}
\begin{split}
&u^{\alpha}(z,\theta)=\sqrt{\frac{\epsilon+\Omega}{2\epsilon}}\left[e^{\tilde{z}}+\frac{\epsilon\cos^2\theta+\Omega\sin^2\theta}{\Omega+i\alpha\Delta\cos\theta}e^{-\tilde{z}}\right]\frac{{\rm sech}\ \tilde{z}}{2},\\
&v_{\parallel}^{\alpha}(z,\theta)=\sin\theta\sqrt{\frac{\epsilon-\Omega}{2\epsilon}}\left[e^{\tilde{z}}+\frac{\Omega}{\Omega+i\alpha\Delta\cos\theta}e^{-\tilde{z}}\right]\frac{{\rm sech}\ \tilde{z}}{2},\\
&v_{\perp}^{\alpha}(z,\theta)=\alpha\cos\theta\sqrt{\frac{\epsilon-\Omega}{2\epsilon}}\left[e^{\tilde{z}}-\frac{\epsilon}{\Omega+i\alpha\Delta\cos\theta}e^{-\tilde{z}}\right]\frac{{\rm sech}\ \tilde{z}}{2},\\
&w^{\alpha}(z,\theta)=\alpha\frac{\sin2\theta}{2}\sqrt{\frac{\epsilon-\Omega}{2\epsilon}}\frac{\Delta}{\Omega+i\alpha\Delta\cos\theta}e^{-\tilde{z}}\frac{{\rm sech}\ \tilde{z}}{2},
\end{split}
\end{equation}
where $\Omega=\sqrt{\epsilon^2-\Delta^2}+i\delta$ has a convergence factor $\delta$.
The analytical expression of the wave function is proper provided that the order of $(\delta/\Delta)^2$ is negligible.
The factor $(\Omega+i\alpha\Delta\cos\theta)^{-1}$ has a pole at $\epsilon=\Delta\sin\theta$ which coincides the energy level of the surface Andreev bound state.
In the limit $z\to+\infty$, the wave function approaches the bulk solutions as $u^{\alpha}\to \sqrt{(\epsilon+\Omega)/2\epsilon}$, $v_{\parallel}^{\alpha}\to \sin\theta\sqrt{(\epsilon-\Omega)/2\epsilon}$, $v_{\perp}^{\alpha}\to \alpha\cos\theta\sqrt{(\epsilon-\Omega)/2\epsilon}$, and $w^{\alpha}\to 0$.

\subsection{Reflection rates of normal reflection and Andreev reflection}

We consider that the particle-like quasiparticle with the up-spin is injected toward the free surface and the injected quasiparticle is reflected at the free surface as the possible four types of quasiparticle.
The injected wave function $\phi^{\rm in}(z,{\bm p}_{\parallel})$ and the reflected wave function $\phi^{\rm re}(z,{\bm p}_{\parallel})$ are described by
\begin{equation}
\begin{split}
\phi^{\rm in}(z,{\bm p}_{\parallel})=&\phi_{\rm p\uparrow}^-(z,{\bm p}_{\parallel}),\\
\phi^{\rm re}(z,{\bm p}_{\parallel})=&r_{\rm N}^{\uparrow\uparrow}(\epsilon,{\bm p}_{\parallel})\phi_{\rm p\uparrow}^+(z,{\bm p}_{\parallel})
+r_{\rm N}^{\uparrow\downarrow}(\epsilon,{\bm p}_{\parallel})\phi_{\rm p\downarrow}^+(z,{\bm p}_{\parallel})\\
&+r_{\rm A}^{\uparrow\uparrow}(\epsilon,{\bm p}_{\parallel})\phi_{\rm h\uparrow}^-(z,{\bm p}_{\parallel})
+r_{\rm A}^{\uparrow\downarrow}(\epsilon,{\bm p}_{\parallel})\phi_{\rm h\downarrow}^-(z,{\bm p}_{\parallel}).
\end{split}
\end{equation}
The coefficients of the reflected wave function give the reflection rates, such as $R_{\rm N(A)}^{\sigma\sigma'}(\epsilon,\theta)=|r_{\rm N(A)}^{\sigma\sigma'}(\epsilon,{\bm p}_{\parallel})|^2$, where the reflection rates are independent of the momentum angle $\phi$ included only in a phase factor of the coefficients. 
The reflection rate $R_{\rm N}^{\uparrow\uparrow}$ ($R_{\rm N}^{\uparrow\downarrow}$)  indicates the normal reflection without (with) the spin flip and $R_{\rm A}^{\uparrow\uparrow}$ ($R_{\rm A}^{\uparrow\downarrow}$) indicates the Andreev reflection without (with) the spin flip.
The coefficients are determined from the boundary condition at $z=0$, $\phi^{\rm in}(0,{\bm p}_{\parallel})+\phi^{\rm re}(0,{\bm p}_{\parallel})=0$, which implies that the superposition wave function vanishes at the free surface.

In the limit of the convergence factor $\delta\to 0$ in $\Omega$, the reflection rates are given by
\begin{equation}
\begin{split}
&R_{\rm N}^{\uparrow\uparrow}(\epsilon,\theta)=\frac{(\epsilon\cos^2\theta+\Omega\sin^2\theta)^2}{\epsilon^2-\Delta^2\sin^2\theta},\
R_{\rm N}^{\uparrow\downarrow}(\epsilon,\theta)=\frac{(\epsilon-\Omega)^2\sin^2\theta\cos^2\theta}{\epsilon^2-\Delta^2\sin^2\theta},\\
&R_{\rm A}^{\uparrow\uparrow}(\epsilon,\theta)=R_{\rm A}^{\uparrow\downarrow}(\epsilon,\theta)=0,
\label{eq:rate1}
\end{split}
\end{equation}
for the quasiparticle energy $\epsilon>\Delta$.
Thus, the Andreev reflection is lacked for the quasiparticles with energy above the bulk energy gap.
The quasiparticle with the energy just equal to the bulk energy gap undergoes the Andreev reflection perfectly.
For the quasiparticle with $\epsilon=\Delta$, the reflection rates become
\begin{equation}
R_{\rm N}^{\uparrow\uparrow}(\Delta,\theta)=R_{\rm N}^{\uparrow\downarrow}(\Delta,\theta)=0,\
R_{\rm A}^{\uparrow\uparrow}(\Delta,\theta)=\sin^2\theta,\
R_{\rm A}^{\uparrow\downarrow}(\Delta,\theta)=\cos^2\theta.
\end{equation}

\begin{figure}
  \begin{center} \resizebox{\textwidth}{!}{%
\includegraphics{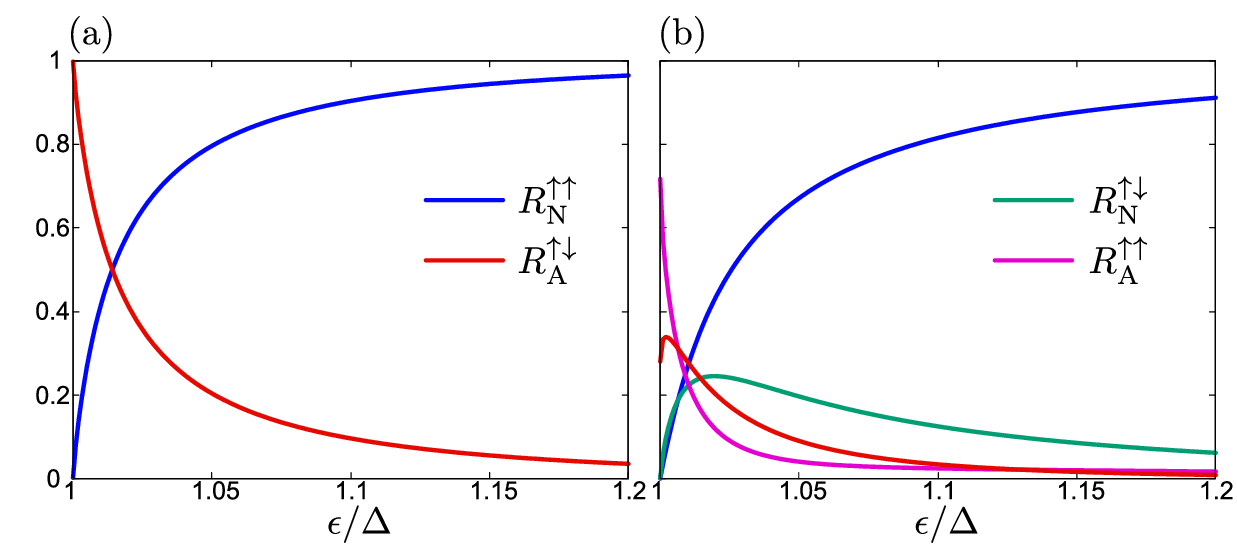}}
\caption{\label{fig:rate_E} (Color online) Quasiparticle energy dependence of the reflection rates for $\theta=0^{\circ}$ (a) and $\theta=60^{\circ}$ (b).
The reflection rates are calculated with the convergence factor $\delta=0.2\Delta$.}
  \end{center}
\end{figure}

A finite convergence factor $\delta$ induces the Andreev reflection for the quasiparticles with the energy near the bulk energy gap.
The quasiparticle energy dependence of the reflection rates for the incident angles $\theta=0^{\circ}$ and $\theta=60^{\circ}$ with $\delta=0.2\Delta$ are shown in Figs.~\ref{fig:rate_E}a and \ref{fig:rate_E}b, respectively.
The Andreev reflection occurs for low energy quasiparticles.
The high energy quasiparticles are dominated by the normal reflection without the spin flip, where the reflection rates approach to the values given by Eq.~\eqref{eq:rate1}.
For $\theta=0^{\circ}$, since the quasiparticles do not feel the parallel component of the pair potential $\Delta_{\parallel}$, the injected particle-like quasiparticles are reflected as the particle-like quasiparticles without the spin flip or the hole-like quasiparticles with the spin flip.
For the finite incident angle, the normal reflection with the spin flip and the Andreev reflection without the spin flip are induced owing to $\Delta_{\parallel}$.
The quasiparticle energy dependence of the reflection rates for these incident angles are qualitatively in agreement with the results by Nagato {\it et al.}~\cite{Nagato1998} which are numerically calculated under a self-consistently determined pair potential with the specular wall.

\begin{figure}
  \begin{center} \resizebox{\textwidth}{!}{%
\includegraphics{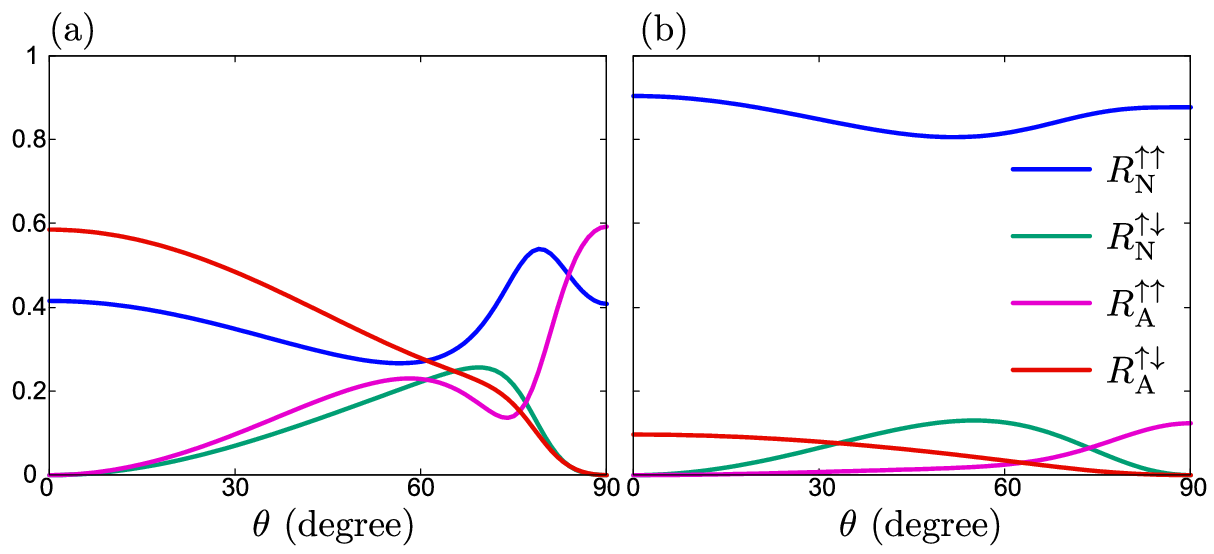}}
\caption{\label{fig:rate_t} (Color online) Incident angle dependence of the reflection rates for the quasiparticle energy $\epsilon=1.01\Delta$ (a) and $\epsilon=1.1\Delta$ (b).
They are calculated with $\delta=0.2\Delta$.}
  \end{center}
\end{figure}

The incident angle dependence of the reflection rates for the quasiparticle energy $\epsilon=1.01\Delta$ and $\epsilon=1.1\Delta$ is shown in Figs.~\ref{fig:rate_t}a and \ref{fig:rate_t}b, respectively.
The rate of the Andreev reflection without the spin flip grows for large incident angles in the vicinity of the bulk energy gap.
For the quasiparticle with energy $\epsilon=1.1\Delta$, in contrast, the Andreev reflection rate is not enhanced even in large incident angles and the normal reflection dominates in every incident angle.
Thus, the Andreev reflection for quasiparticles with large incident angles is strongly suppressed in high energy by comparison to the result in Ref.~\cite{Nagato1998}.
This difference may arise from the spatial variation of $\Delta_{\parallel}$ obtained by the self-consistent calculation.

\begin{figure}
  \begin{center} \resizebox{0.7\textwidth}{!}{%
\includegraphics{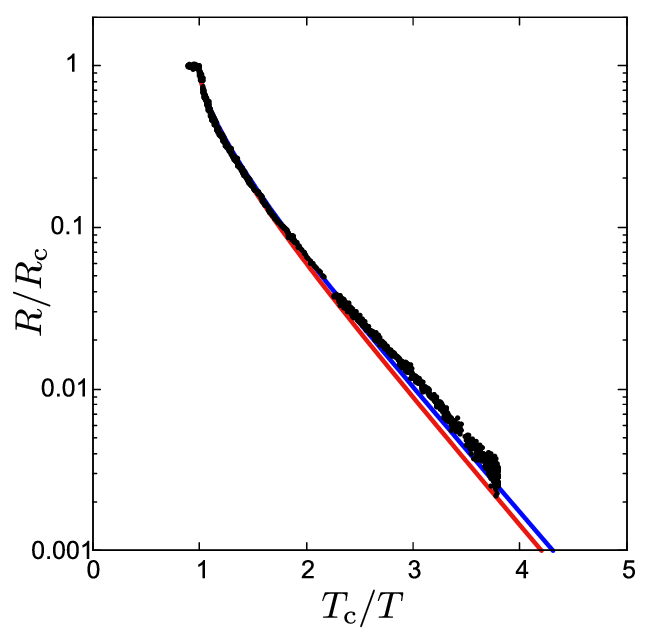}}
\caption{\label{fig:resistivity} (Color online) 
Temperature dependence of the resistivity of the Wigner solid normalized by $R_{\rm c}\equiv R(T_{\rm c})$.
A small magnetic field of 0.07~T is applied for experimental data (solid circle), but no essential difference is observed under no magnetic field.
Blue line indicates the theoretically obtained resistivity while taking into account the contribution of the Andreev reflection by the quasiparticle with $\epsilon=\Delta$ in the limit $\delta\to 0$.
Red line is calculated from Eq.~\eqref{eq:normalized-mobility-2} with reflection rates for $\delta=0.2\Delta$.
The theoretical curves overlap behind the experimental data near $T_{\rm c}$.}
  \end{center}
\end{figure}

\subsection{Mobility of Wigner solid}

The mobility of the Wigner solid can be calculated from Eq.~\eqref{eq:normalized-mobility-2} with $T_{00}(\epsilon,\theta)=R_{\rm N}^{\uparrow\uparrow}(\epsilon,\theta)+R_{\rm N}^{\uparrow\downarrow}(\epsilon,\theta)$.
In the limit $\delta\to 0$, the only quasiparticle with the energy $\epsilon(\eta_p)=\Delta$, namely $\eta_p=0$, undergoes the Andreev reflection with $T_{00}=0$.
Since the factor $\partial \epsilon/\partial\eta_p=\eta_p/\epsilon=0$ in Eq.~\eqref{eq:normalized-mobility-2} for the Andreev reflected quasiparticle, we can regard $T_{00}$ as unity in the calculation.
Then, the reciprocal of the normalized mobility $\mu^{\rm n}/\mu$ in Eq.~\eqref{eq:normalized-mobility-2} is equivalent to the normalized resistivity $R/R_{\rm c}$ in Eq.~\eqref{eq:fermi-function} with $R_{\rm c}\equiv R(T_{\rm c})$.
The temperature dependence of the resistivity is shown in Fig.~\ref{fig:resistivity} by the blue line with the experimental data (solid circle)~\cite{Ikegami2006}, where the bulk energy gap $\Delta(T)$ is calculated by the BCS gap equation.
The theoretical result is in good agreement with the experimental observations, where the theoretical curve overlaps behind the experimental data near $T_{\rm c}$.
We also plot the resistivity with the finite convergence factor $\delta=0.2\Delta$ by the red line in order to clarify the suppression of the resistivity by the Andreev reflection.
The resistivity of the Wigner solid is suppressed in low temperatures due to the Andreev reflection of the low energy quasiparticles near the bulk energy gap.
The resistivity tends to be strongly suppressed for large $\delta$.

\section{Discussion}\label{sec4}

The Andreev reflection is lacked for quasiparticles with $\epsilon>\Delta$ under the present model potential $\Delta_{\parallel}(z)=\Delta$ and $\Delta_{\perp}(z)=\Delta\tanh\tilde{z}$ in the limit of the convergence factor $\delta\to 0$.
This result is qualitatively different under the uniform model potential $\Delta_{\parallel}(z)=\Delta_{\perp}(z)=\Delta$ in which the finite Andreev reflection rate $R_{\rm A}^{\uparrow\uparrow}+R_{\rm A}^{\uparrow\downarrow}=\Delta^2\cos^2\theta/(\epsilon^2-\Delta^2\sin^2\theta)$ is obtained without the convergence factor.
The convergence factor $\delta$ gives the spatial decay of the quasiparticle wave function with the characteristic length $(\Delta/\delta)\xi$.
The present result implies that the low energy quasiparticles loosely bound near the free surface only undergo the Andreev reflection.

The parallel component of the order parameter $\Delta_{\parallel}(z)$ increases near the free surface according to the self-consistent calculation~\cite{Nagato1998}.
The quasiparticles with the incident angle $\theta$ feel the pair potential $\Delta_{\parallel}(0)\sin\theta$ at the free surface.
The incident quasiparticles nearly parallel to the free surface within the energy range $\epsilon<\Delta_{\parallel}(0)\sin\theta$ will undergo the Andreev reflection.
However, the Andreev reflection with the large incident angle hardly contributes to a rise in the mobility of the Wigner solid due to the factor $\cos^3\theta$ in Eq.~\eqref{eq:normalized-mobility-2}.

\section{Conclusion}\label{sec5}

We have analytically calculated the contribution of the Andreev retroreflection to the mobility of the Wigner solid on superfluid $^3$He-B by employing a realistic model order parameter, $\Delta_{\parallel}(z)=\Delta$ and $\Delta_{\perp}(z)=\Delta\tanh(z/\xi)$.
We have found analytical solutions of the Andreev equation giving the quasiparticle wave function.
The Andreev retroreflection is lacked for quasiparticles with energy above the bulk energy gap under the model order parameter.
Then, the Andreev retroreflection does not contribute to a rise in the mobility of the Wigner solid on the superfluid $^3$He-B.
The present model calculation is in good agreement with the previous experimental observation.
Note that the parallel component of the order parameter $\Delta_{\parallel}(z)$ increases near the free surface according to the self-consistent calculation.
The increase in $\Delta_{\parallel}(z)$ will cause the Andreev retroreflection for the incident quasiparticles nearly parallel to the free surface.
Quantitative estimation of the contribution of this Andreev retroreflection to the mobility remains as future work.


\backmatter

\bmhead{Acknowledgments} We are grateful to Y. Nagato and K. Nagai for
discussions in the early stage of this research.

\bmhead{Author Contributions} Y.T. carried out theoretical calculations.
All authors equally contributed to the manuscript preparation.

\bmhead{Funding} This work is supported by National Science and
Technology Council, Taiwan (grant No. NSTC 112-2112-M-A49-040-), the
Higher Education Sprout Project by the Ministry of Education (MOE) in
Taiwan, and JSPS KAKENHI Grant No. 21K03456.

\bmhead{Data Availability} The datasets generated during the current
study are available from the corresponding author on reasonable
request.

%

\bibliography{reference}

\end{document}